\documentclass[twocolumn,description,aps]{revtex4-1}
\usepackage{graphicx,epstopdf,amsmath,amssymb,multirow,CJK,color}
\usepackage[german,english]{babel}

\begin{document}

\title{Triply degenerate nodal points in RERh$_{6}$Ge$_{4}$ (RE=Y, La, Lu)}

\author{Peng-Jie Guo}
\author{Huan-Cheng Yang}
\author{Kai Liu}\email{kliu@ruc.edu.cn}
\author{Zhong-Yi Lu}\email{zlu@ruc.edu.cn}

\affiliation{Department of Physics and Beijing Key Laboratory of Opto-electronic Functional Materials $\&$ Micro-nano Devices, Renmin University of China, Beijing 100872, China}
\date{\today}

\begin{abstract}
As a new type of fermions without counterpart in high energy physics, triply degenerate fermions show exotic physical properties, which are represented by triply degenerate nodal points in topological semimetals. Here, based on the space group theory analysis, we propose a practical guidance for seeking a topological semimetal with triply degenerate nodal points located at a symmetric axis, which is applicable to both symmorphic and nonsymmorphic crystals. By using this guidance in combination with the first-principles electronic structure calculations, we predict a class of triply degenerate topological semimetals RERh$_{6}$Ge$_{4}$ (RE=Y, La, Lu). In these compounds, the triply degenerate nodal points are located at the $\Gamma$-A axis and not far from the Fermi level. Especially, LaRh$_{6}$Ge$_{4}$ has a pair of triply degenerate nodal points located very closely to the Fermi level. Considering the fact that the single crystals of RERh$_{6}$Ge$_{4}$ have been synthesized experimentally, the RERh$_{6}$Ge$_{4}$ class of compounds will be an appropriate platform for studying exotic physical properties of triply degenerate topological semimetals.
\end{abstract}

\date{\today} \maketitle

\section{INTRODUCTION}

The topological properties of materials have attracted intensive attention both theoretically and experimentally in recent years~\cite{Hasan, Qi}. With the in-depth study, topological materials have been naturally extended from insulators to semimetals~\cite{Fang-2016}. In comparison with Dirac and Weyl fermions in high energy physics that are necessarily restricted by Lorentz invariance, Dirac points in condensed matter physics are protected by time-reversal, space-inversion, and certain crystal symmetries, while Weyl points can be obtained by breaking either time-reversal or space-inversion symmetry for Dirac semimetal. To date, both Dirac and Weyl points in realistic materials have been predicted by theories~\cite{Wang-2012, Wang-2013, Weng-2015} and then verified by experiments~\cite{Liu-science, Liu-NM, Xu-science, Lv-PRX}. Due to the abundant crystal symmetries relative to Lorentz invariance, it is a natural idea to search new types of fermions in condensed matter physics without counterparts in high energy physics.

Very recently, a new type of fermions without counterpart in high energy physics is triply degenerate fermion, represented by triply degenerate nodal point in topological semimetal, which was first proposed to exist at a high symmetry point in the Brillouin zone and protected by nonsymmorphic symmetry~\cite{Bradlyn-science} and then suggested to also exist at a symmetric axis~\cite{Zhu-PRX, Weng-PRB, Weng-ZrTe, Chang-2017, Yan-PRL, Wang-CuTe}. Triply degenerate nodal point is considered as an intermediate state between doubly-degenerate Weyl point and fourfold-degenerate Dirac point, and it has been observed in several compounds such as MoP and WC~\cite{Lv-nature, Ma-arxiv}. In comparison with Dirac and Weyl semimetals, triply degenerate topological semimetals show exotic properties such as helical anomaly and Lifshitz transition of Fermi surface~\cite{Weng-PRB}. Here, a topological semimetal with triply degenerate nodal points located closely to the Fermi level is very crucial for studying those exotic properties in experiment. However, such a topological semimetal is still lacking. Thus, it is very urgent to search for a new topological semimetal with triply degenerate nodal points close enough to the Fermi level.

In this paper, we propose a practical guidance for searching a triply degenerate topological semimetal, based on the space group theory analysis. By combining this guidance with the first-principles electronic structure calculations, we predict that the RERh$_{6}$Ge$_{4}$ (RE=Y, La, Lu) class of compounds own triply degenerate nodal points whose energies, especially in LaRh$_{6}$Ge$_{4}$, are very close to the Fermi level.

\section{COMPUTATIONAL DETAILS}
\label{sec:Method}

First-principles electronic structure calculations were performed with the projector augmented wave (PAW) method~\cite{B-PRB, Kresse-PRB} as implemented in VASP package~\cite{Kresse-1993, Kresse-1996, Kresse-J}. The generalized gradient approximation (GGA) of Perdew-Burke-Ernzerhof (PBE) type was adopted for the exchange-correlation functional~\cite{Perdew-PRL}. The kinetic energy cutoff of the plane wave basis was set to 400 eV. A 6$\times$6$\times$10 $k$-point mesh for Brillouin zone (BZ) sampling and the Gaussian smearing method with a width of 0.05 eV around the Fermi surface were adopted. Both cell parameters and internal atomic positions were fully relaxed until the forces on all atoms were smaller than 0.01 eV/{\AA}. Once the equilibrium structures were obtained, the electronic structures were calculated by including the spin-orbit coupling (SOC). The maximally localized Wannier function (MLWF) method was used to calculate the Fermi surface~\cite{Marzari-PRB, Marzari-2001}.

\begin{figure*}[!t]
\centering
\includegraphics[width=0.8\textwidth]{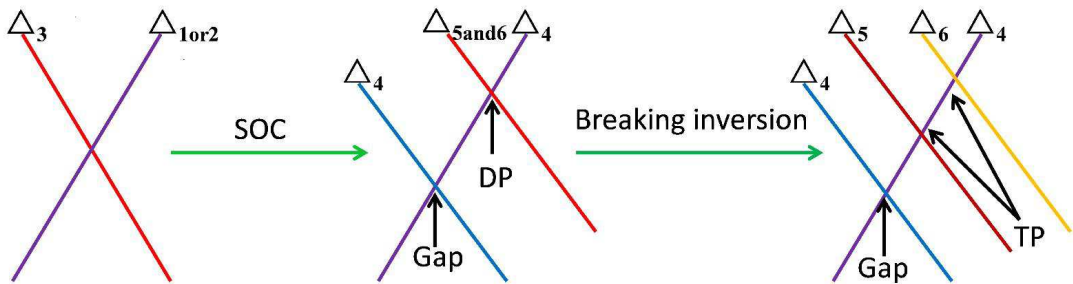}
\caption{(Color online) Schematic illustration for generation of triply degenerate nodal point. The crystal contains time-reversal symmetry, space-inversion symmetry, and C$_{3v}$ group symmetry along a certain direction. The $\Delta_{i}$ labels the irreducible representation in the C$_{3v}$ group and C$_{3v}$ double group. The 'DP' denotes dirac point and the 'TP' represents triply degenerate nodal point. }
\label{Fig1}
\end{figure*}

\section{RESULTS AND ANALYSIS}
\label{sec:Results}

 Here we would like to present a practical guidance for searching a topological semimetal with triply degenerate nodal points, based on the space group theory analysis. We consider a crystal, being either symmorphic or nonsymmorphic, with time-reversal symmetry and space-inversion symmetry as well as the C$_{3v}$ group symmetry along a certain direction, in which there are two bands crossing with each other. It is known that the C$_{3v}$ group contains two one-dimensional ($\Delta_{1}$ and $\Delta_{2}$) and one two-dimensional ($\Delta_{3}$) irreducible representations respectively (see the Appendix). If the two crossing bands belong to the $\Delta_{1or2}$ and $\Delta_{3}$ representations (left panel of Fig. 1) respectively, their crossing point will be a threefold degenerate point, protected by the C$_{3v}$ group symmetry when the spin degree of freedom is ignored. Once the SOC is included, the crystal symmetry transforms the C$_{3v}$ group symmetry to the C$_{3v}$ double group symmetry, which now has four one-dimensional ($\Delta_{1}$, $\Delta_{2}$, $\Delta_{5}$, and $\Delta_{6}$) and two two-dimensional ($\Delta_{3}$ and $\Delta_{4}$) irreducible representations respectively (see the Appendix). The calculation of the character further shows that both $\Delta_{1}$ and $\Delta_{2}$ representations in the C$_{3v}$ group change to the $\Delta_{4}$ representation in the C$_{3v}$ double group, while the $\Delta_{3}$ representation in the C$_{3v}$ group is split into the $\Delta_{4}$, $\Delta_{5}$, and $\Delta_{6}$ representations in the C$_{3v}$ double group (middle panel of Fig. 1) (see the Appendix). Since the crystal has both time-reversal and space-inversion symmetries, the two bands which belong respectively to the $\Delta_{5}$ and $\Delta_{6}$ representations must be degenerate to form the Kramer doublets. Thus the band crossing point that derives from  the $\Delta_{4}$, $\Delta_{5}$, and $\Delta_{6}$ representations will be a Dirac point (DP in the middle panel of Fig. 1), which is protected by time-reversal, space-inversion, and the C$_{3v}$ double group symmetries, similar to the case in the PtSe$_{2}$ class~\cite{Duan-PRB}. On the other hand, the two bands belonging to the same $\Delta_{4}$ representation will hybridize with each other and then open a band gap due to the band repulsion ( the lower crossing in the middle panel of Fig. 1). Next, if the space-inversion symmetry of the crystal is broken, the twofold degenerate band belonging to the $\Delta_{5}$ and $\Delta_{6}$ representations will be split into two non-degenerate bands. Both of them cross with the doubly degenerate $\Delta_{4}$ band and then give rise to a pair of triply degenerate nodal points (TP in the right panel of Fig. 1), which are protected by time-reversal and the C$_{3v}$ double group symmetries but without space-inversion symmetry. Actually, the further calculation and analysis of the characters demonstrate that except the C$_{3v}$ group, all the other point groups cannot provide such a protection. As a result, triply degenerate nodal points located at a symmetric axis only likely exist in a system with time-reversal and the C$_{3v}$ double group symmetries but without space-inversion symmetry. Just as described, here it should be emphasized that we can apply this guidance to both symmorphic and nonsymmorphic crystals.

\begin{figure}[!b]
\includegraphics[width=0.95\columnwidth]{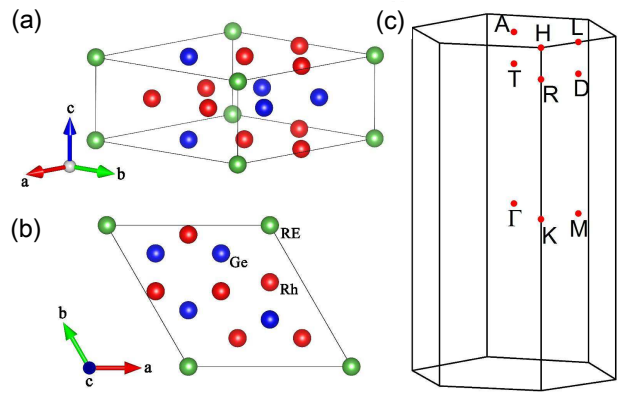}
\caption{(Color online) (a) Perspective view and (b) top view of the crystal structure of RERh$_{6}$Ge$_{4}$ (RE=Y, La, Lu). (c) Bulk Brillouin zone. Red dots label the high-symmetry $\textit{k}$ points.}
\label{Fig2}
\end{figure}

 Now let us examine ternary compounds RERh$_{6}$Ge$_{4}$ (RE=Y, La, Lu) which own the P$\overline{6}$m2 space group symmetry and the corresponding D$_{3h}$ point group symmetry. The perspective and top views for the unit cell of RERh$_{6}$Ge$_{4}$ crystal are shown in Figs. 2(a) and 2(b), respectively. As we see, there is the C$_{3v}$ group symmetry along the c-axis but without space-inversion symmetry. Moreover, the previous experimental study indicates that the compounds RERh$_{6}$Ge$_{4}$ are nonmagnetic and possess time-reversal symmetry~\cite{LaRhGe}. Thus, according to the above guidance, the compounds RERh$_{6}$Ge$_{4}$ are such candidate materials with triply degenerate nodal points along the $\Gamma$-A ($k_c$) direction in the BZ [Fig. 2(c)].

\begin{figure}[!t]
\includegraphics[width=0.8\columnwidth]{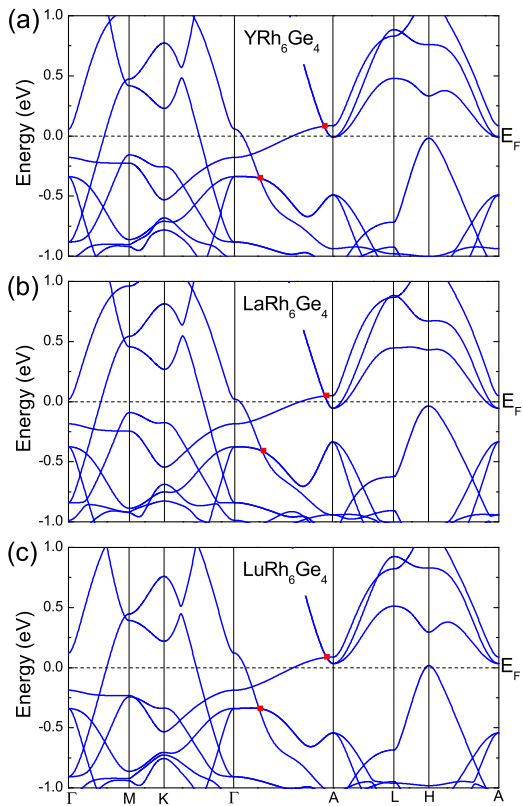}
\caption{(Color online) Band structures of (a) YRh$_{6}$Ge$_{4}$, (b) LaRh$_{6}$Ge$_{4}$, and (c) LuRh$_{6}$Ge$_{4}$ along high symmetry paths in the BZ calculated without the SOC. The triply degenerate nodal points are marked by red dots.}
\label{Fig3}
\end{figure}

Following the schematic illustration for generation of triply degenerate nodal points (Fig. 1), we first calculate the band structures of RERh$_{6}$Ge$_{4}$ (RE=Y, La, Lu) without the SOC. As shown in Fig. 3, the band structures of YRh$_{6}$Ge$_{4}$, LaRh$_{6}$Ge$_{4}$, and LuRh$_{6}$Ge$_{4}$ are very similar. They all have two pairs of threefold degenerate points (indicated by red dots) around the Fermi level $E_F$ along the $\Gamma$-A direction in the BZ. These threefold degenerate points are generated by the crossing of a doubly degenerated band and a nondegenerate band, and exhibit the characteristic described in the left panel of Fig. 1. Importantly, the upper threefold degenerate point is very close to the Fermi level, and it may move below the $E_F$ when considering the SOC. In addition, they all have nodal rings around the $\Gamma$ point at about -0.25 eV, which are protected by the mirror reflection symmetry. Since we focus on the triply degenerate nodal point, the nodal ring will not be discussed further.

\begin{figure}[!t]
\includegraphics[width=0.8\columnwidth]{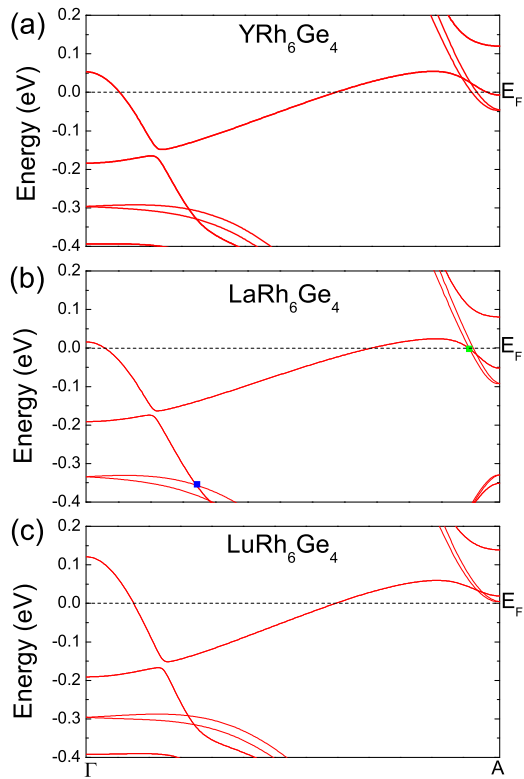}
\caption{(Color online) Band structures of (a) YRh$_{6}$Ge$_{4}$, (b) LaRh$_{6}$Ge$_{4}$, and (c) LuRh$_{6}$Ge$_{4}$ along the $\Gamma$-A direction calculated with the SOC. Two pairs of triply degenerate nodal points in LaRh$_{6}$Ge$_{4}$ are located at the green and blue dots, respectively.}
\label{Fig4}
\end{figure}

 Figure 4 shows the band structures of RERh$_{6}$Ge$_{4}$ along the $\Gamma$-A direction calculated with the SOC. By comparing Fig. 3 and Fig. 4, we see that each threefold degenerate point without the SOC indeed changes to a pair of triply degenerate nodal points when including the SOC, as demonstrated in Fig. 1. Although the upper pairs of triply degenerate nodal points for YRh$_{6}$Ge$_{4}$ and LuRh$_{6}$Ge$_{4}$ [Figs. 4(a) and 4(c)] are still above the Fermi level, the corresponding pair of triply degenerate nodal points for LaRh$_{6}$Ge$_{4}$ locate very close to the Fermi level [green dot in Fig. 4(b)] at the T point along the $\Gamma$-A axis in the BZ [Fig. 2(c)]. Moreover, the other pair of triply degenerate nodal points lie around 300-400 meV below the Fermi level [blue dot in Fig. 4(b)]. These triply degenerate nodal points along the $\Gamma$-A direction should be observed in the angle-resolved photoemission spectroscopy (ARPES) experiment. Hereafter, we focus on the LaRh$_{6}$Ge$_{4}$ compound.

\begin{figure}[!t]
\includegraphics[width=0.8\columnwidth]{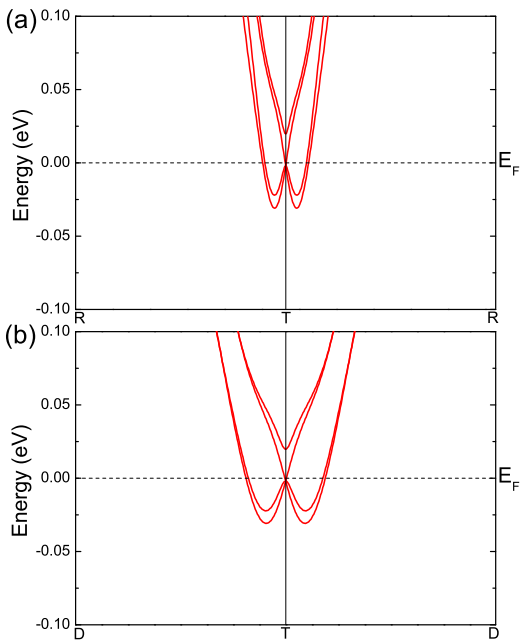}
\caption{(Color online) Band structure of LaRh$_{6}$Ge$_{4}$ for the four bands around the triply degenerate nodal point [green dot in Fig. 4(b)] calculated with the SOC along (a) R-T-R and (b) D-T-D directions in the BZ.}
\label{Fig5}
\end{figure}

It is known that different directions in the BZ may have different symmetries. For LaRh$_{6}$Ge$_{4}$, the $\Gamma$-A direction owns the highest symmetry among various directions in the BZ. A twofold degenerate band along the $\Gamma$-A direction may become nondegenerated along another direction in the BZ; likewise, a linear band-crossing around a triply degenerate nodal point along the $\Gamma$-A direction may become nonlinear along another direction of the BZ. Here, Figs. 5(a) and 5(b) show the band structures of LaRh$_{6}$Ge$_{4}$ around the upper triply degenerate nodal point (indicated by green dot in Fig. 4(b)) along the R-T-R and D-T-D directions [Fig. 2(c)], respectively. The band dispersions along these two directions are very similar, but distinct from that along the $\Gamma$-A direction. The triply degenerate nodal point is generated by the band-crossing of two linear nondegenerate bands with one parabolic band along the R-T-R and D-T-D directions, demonstrating the anisotropy of bands due to different symmetries along different directions.

\begin{figure}[!t]
\includegraphics[width=0.96\columnwidth]{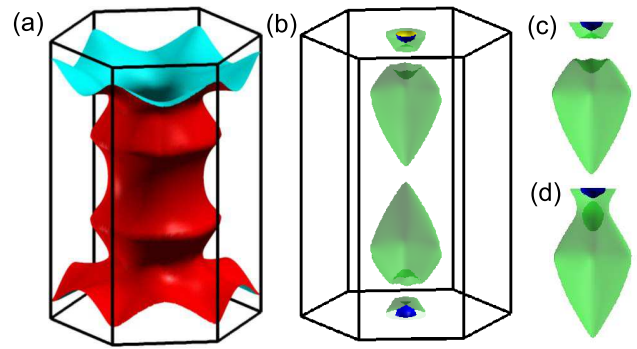}
\caption{(Color online)  (a) The hole-type and (b) the electron-type Fermi surface sheets of LaRh$_{6}$Ge$_{4}$ at 10 meV below the Fermi level. Green and blue isosurfaces represent two different electron-type Fermi surface sheets. Enlarged upper electron-type Fermi surface sheets of LaRh$_6$Ge$_4$ at (c) -10 meV and (d) 0 meV with respect to the Fermi level, respectively.}
\label{Fig6}
\end{figure}

In type-I Dirac or Weyl semimetal, the Fermi surface consists of discrete points when the Fermi level is shifted to the energy of Dirac or Weyl point. In comparison, the Fermi surface of topological semimetal with triply degenerate nodal point is rather complex. For LaRh$_{6}$Ge$_{4}$, the Fermi surface sheets at 10 meV below the E$_{F}$ are shown in Fig. 6. Among the six Fermi surface sheets, three sheets are very similar to the other three ones, thus we only show one hole-type [Fig. 6(a)] and two electron-type [Fig. 6(b)] Fermi surface sheets here. As shown in Fig. 6(b), the two electron-type pockets have a single connective point along the $\Gamma$-A direction. Actually, the touching of the two electron-type Fermi pockets exists in a range of chemical potentials [Fig 6 (c)], resulting from the properties of topological semimetal with triply degenerate nodal point~\cite{Weng-PRB}. This is also distinct from type-II Dirac or Weyl semimetal, in which the hole-type and electron-type Fermi surface sheets touch each other only when the Fermi level crosses the type-II Dirac or Weyl point~\cite{Dai-nature}. By tuning the Fermi level of LaRh$_{6}$Ge$_{4}$ out of the energy range between the two triply degenerate nodal points, the two connective Fermi surface sheets separate from each other, inducing a Lifshitz transition [Fig 6 (c) and 6(d)].

\section{DISCUSSION AND SUMMARY}
\label{sec:discussion}

Symmetries play an important role in physics studies. For a crystal with time-reversal symmetry, space-inversion symmetry, and the C$_{3v}$ symmetry, the analysis based on the space group theory indicates that if the crystal has a threefold degenerate point along an axis with the C$_{3v}$ group symmetry when the SOC is ignored (left panel of Fig. 1), a Dirac point will take place when including the SOC (middle panel of Fig. 1). Further, if space-inversion symmetry is broken, the Dirac point will become a pair of triply degenerate nodal points (right panel of Fig. 1). Thus, if one attempts to find Dirac or triply degenerate nodal points, one only needs to perform a simple calculation without the SOC on a crystal satisfying the above symmetries and then examine whether it contains threefold degenerate points along a direction with the C$_{3v}$ group symmetry. This provides us a practical guidance to effectively search a topological semimetal with the triply degenerate nodal points located at a symmetric axis.

With the help of the guidance, we seek out the RERh$_{6}$Ge$_{4}$ (RE=Y, La, Lu) class as material examples for topological semimetals with triply degenerate nodal points. These compounds have many prominent characteristics. First, the energies of their triply degenerate nodal points around the A point in the BZ are within a range of 50-meV from the Fermi level, in contrast with the 180-meV value in previously found topological semimetals such as WC~\cite{Zhu-PRX, Chang-2017, Ma-arxiv} . Especially, a pair of triply degenerate nodal points in LaRh$_{6}$Ge$_{4}$ locate just 3 meV below the $E_F$ [Fig. 4(b)], which may readily show the transport anomalous properties. Second, the single crystals of the RERh$_{6}$Ge$_{4}$ class have already been synthesized~\cite{LaRhGe}, which will facilitate the experimental study on their exotic physical properties. Third, the bands of triply degenerate nodal points around the A point in the RERh$_{6}$Ge$_{4}$ class are all of electron-type. Thus, by tuning the Fermi level around the triply degenerate nodal points, the corresponding Fermi surfaces are also all of electron-type, which is distinct from the other topological semimetals with triply degenerate nodal points, such as TaN~\cite{Weng-PRB}. Fourth, since the touching point of Fermi surfaces in the RERh$_{6}$Ge$_{4}$ class is protected by the C$_{3}$ symmetry (Fig. 6), the 'magnetic breakdown'~\cite{Weng-PRB} may also be observed in quantum oscillation measurement. Furthermore, compared with triply degenerate nodal points located at high symmetry points in the BZ, a triply degenerate nodal point located at a symmetric axis can be moved along the axis by tuning parameters and is thus more robust.

%\section{SUMMARY}
%\label{sec:summary}

 In summary, based on the space group theory analysis, we propose a practical guidance to effectively seek triply degenerate nodal points in materials with time-reversal symmetry and the C$_{3v}$ group symmetry along a certain direction but without space-inversion symmetry, which is applicable to both symmorphic and nonsymmorphic crystals. Accordingly, we predict a class of topological semimetals with triply degenerate nodal points: RERh$_{6}$Ge$_{4}$ (RE=Y, La, Lu), which provide an appropriate platform for studying exotic physical properties of triply degenerate topological semimetal, such as the Fermi arcs, transport anomalies, and topological Lifshitz transitions.

\begin{acknowledgments}

We thank Z. X. Liu for helpful conversations. This work was supported by National Key R\&D Program of China (Grant No. 2017YFA0302903), National Natural Science Foundation of China (Grants No. 11474356, No. 91421304, and No. 11774424), the Fundamental Research Funds for the Central Universities, and the Research Funds of Renmin University of China (Grants No. 14XNLQ03 and No. 16XNLQ01). Computational resources have been provided by the Physical Laboratory of High Performance Computing at Renmin University of China.

\begin{appendix}
\section{symmetry analysis }
\label{sec:appendix}

We show the character tables of C$_{3v}$ and C$_{3v}^{d}$ groups in Tables I and II, respectively.

\begin{table}[b!]
\caption{\label{tab:I} Character table of the group C$_{3v}$.}
\begin{center}
\begin{tabular*}{1.0\columnwidth}{@{\extracolsep{\fill}}cccc}
\hline\hline
       C$_{3v}$ &      E&      2C$_{3}$  &    3$\sigma_{v}$ \\
\hline
  $\Delta_{1}$ &	1 &  1 & 1  \\
  $\Delta_{2}$ &    1 &  1 & -1  \\
  $\Delta_{3}$ &    2 & -1 & 0   \\
\hline\hline
\end{tabular*}
\end{center}
\end{table}

\begin{table}[b!]
\caption{\label{tab:II}Character table of the double group C$_{3v}^{d}$. The 'e' represents the symmetry operation of 2$\pi$ rotation.}
\begin{center}
\begin{tabular*}{1.0\columnwidth}{@{\extracolsep{\fill}}ccccccc}
\hline\hline
       C$_{3v}^{d}$ &      E&      e  &   (C$_{3}$, eC$_{3}^{2}$)&  (C$_{3}^{2}$, eC$_{3}$)& 3$\sigma_{v}$ & 3e$\sigma_{v}$ \\
\hline
  $\Delta_{1}$ &	1 &  1 & 1& 1& 1& 1 \\
  $\Delta_{2}$ &    1 &  1 & 1& 1& -1&-1 \\
  $\Delta_{3}$ &    2 & 2 & -1& -1& 0& 0  \\
  $\Delta_{4}$ &    2 & -2 & 1& -1& 0& 0  \\
  $\Delta_{5}$ &    1 & -1 & -1& 1& i& -i  \\
  $\Delta_{6}$ &    1 & -1 & -1& 1& -i& i  \\
\hline\hline
\end{tabular*}
\end{center}
\end{table}

\begin{table}[b!]
\caption{\label{tab:III} The character of the $\Delta_{1}$ representation with the SOC.}
\begin{center}
\begin{tabular*}{1.0\columnwidth}{@{\extracolsep{\fill}}ccccccc}
\hline\hline
    $\Delta_{1}$    &      E&      e  &   (C$_{3}$, eC$_{3}^{2}$)&  (C$_{3}^{2}$, eC$_{3}$)& 3$\sigma_{v}$& 3e$\sigma_{v}$ \\
\hline
  SOC &    2 & -2 & 1& -1& 0& 0  \\
\hline\hline
\end{tabular*}
\end{center}
\end{table}

\begin{table}[b!]
\caption{\label{tab:IV} The character of the $\Delta_{2}$ representation with the SOC.}
\begin{center}
\begin{tabular*}{1.0\columnwidth}{@{\extracolsep{\fill}}ccccccc}
\hline\hline
    $\Delta_{2}$    &      E&      e  &   (C$_{3}$, eC$_{3}^{2}$)&  (C$_{3}^{2}$, eC$_{3}$)& 3$\sigma_{v}$& 3e$\sigma_{v}$ \\
\hline
  SOC &    2 & -2 & 1& -1& 0& 0  \\
\hline\hline
\end{tabular*}
\end{center}
\end{table}

\begin{table}[b!]
\caption{\label{tab:V} The character of the $\Delta_{3}$ representation with the SOC.}
\begin{center}
\begin{tabular*}{1.0\columnwidth}{@{\extracolsep{\fill}}ccccccc}
\hline\hline
    $\Delta_{3}$    &      E&      e  &   (C$_{3}$, eC$_{3}^{2}$)&  (C$_{3}^{2}$, eC$_{3}$)& 3$\sigma_{v}$& 3e$\sigma_{v}$ \\
\hline
  SOC &    4 & -4 & -1& 1& 0& 0  \\
\hline\hline
\end{tabular*}
\end{center}
\end{table}

The calculations indicate that both the irreducible representations $\Delta_{1}$  and $\Delta_{2}$ in the C$_{3v}$ group change to the irreducible representation $\Delta_{4}$ of the C$_{3v}$ double group (Tables III and IV) when including the SOC. While the irreducible representation $\Delta_{3}$ in the C$_{3v}$ group changes to a reducible representation of the C$_{3v}$ double group (Table V). In accordance with the character of the $\Delta_{3}$ reducible representation and the character table of the C$_{3v}$ double group in Table II, the $\Delta_{3}$ reducible representation can be reduced by the formula of reduced coefficient as following,

\begin{equation*}
a_j = \frac{1}{g}\sum_{R\in G}\chi^{j}(R)\ast\chi(R)
\end{equation*}

Here, the reduced coefficient $a_{j}$ represents the number of appearance for irreducible representation $j$ in a reducible representation, $g$ is the number of symmetry operations in group $G$, $\chi^{j }(R)$ is the character of the symmetry operation $R$ in the irreducible representation $j$, $\chi(R)$ is the character of the symmetry operation $R$ in a reducible representation. Specifically, we have

$a_{4}$ = $\frac{1}{12}$[4$\ast$2+(-4)$\ast$(-2)+2$\ast$(-1)$\ast$1+2$\ast$(-1)$\ast$1+0+0]=1,

$a_{5}$ = $\frac{1}{12}$[4$\ast$1+(-4)$\ast$(-1)+2$\ast$(-1)$\ast$(-1)+2$\ast$1$\ast$1+0+0]=1,

$a_{6}$ = $\frac{1}{12}$[4$\ast$1+(-4)$\ast$(-1)+2$\ast$(-1)$\ast$(-1)+2$\ast$1$\ast$1+0+0]=1.

%\begin{equation*}
%$a_{4} = \frac{1}{12}[4\ast2+(-4)\ast(-2)+2\ast(-1)\ast1+2\ast(-1)\ast1+0+0]=1$
%\end{equation*}
%\begin{equation*}
%$a_{5} = \frac{1}{12}[4\ast1+(-4)\ast(-1)+2\ast(-1)\ast(-1)+2\ast1\ast1+0+0]=1$
%\end{equation*}
%\begin{equation*}
%$a_{6} = \frac{1}{12}[4\ast1+(-4)\ast(-1)+2\ast(-1)\ast(-1)+2\ast1\ast1+0+0]=1$
%\end{equation*}

Thus, the $\Delta_3$ representation in the C$_{3v}$ group is split into the $\Delta_4$, $\Delta_5$, and $\Delta_6$ representations in the C$_{3v}$ double group.

\end{appendix}

\end{acknowledgments}

\end{document}